\documentclass{JHEP}

\usepackage{epsfig,latexsym}

\def\bef{\begin{figure}}
\def\eef{\end{figure}}

\newcommand{\be}[1]{\begin{equation}\label{#1}}
\newcommand{\beq}{\begin{equation}}
\newcommand{\ee}{\end{equation}}
\newcommand{\beqn}[1]{\begin{eqnarray}\label{#1}}
\newcommand{\eeqn}{\end{eqnarray}}
\newcommand{\bd}{\begin{displaymath}}
\newcommand{\ed}{\end{displaymath}}

\def\lsim{\raise0.3ex\hbox{$\;<$\kern-0.75em\raise-1.1ex
e\hbox{$\sim\;$}}}
\def\gsim{\raise0.3ex\hbox{$\;>$\kern-0.75em\raise-1.1ex
\hbox{$\sim\;$}}}
\def\simlt{\mathrel{\lower2.5pt\vbox{\lineskip=0pt\baselineskip=0pt
           \hbox{$<$}\hbox{$\sim$}}}}
\def\simgt{\mathrel{\lower2.5pt\vbox{\lineskip=0pt\baselineskip=0pt
           \hbox{$>$}\hbox{$\sim$}}}}
\def\unity{{\hbox{1\kern-.8mm l}}}
\newcommand{\ov}{\overline}
\renewcommand{\to}{\rightarrow}

\newcommand{\mc}{\mathcal}

\renewcommand{\to}{\rightarrow}

\newcommand{\br}{\langle}

\newcommand{\kt}{\rangle}

\def\lapprox{\mathrel{\mathop  {\hbox{\lower0.5ex\hbox{$\sim$}
\kern-1.1em\lower-0.7ex\hbox{$<$}}}}}
\def\gapprox{\mathrel{\mathop  {\hbox{\lower0.5ex\hbox{$\sim$}
\kern-1.1em\lower-0.7ex\hbox{$>$}}}}}

\preprint{
  Aug 2004\\
}

\title{Mirror World, Supersymmetric Axion and Gamma Ray Bursts}

\author{%
L. Gianfagna\rlap{$^a$}, M. Giannotti\rlap{$^b$}, F. Nesti\rlap{$^a$}\\
\noindent\llap{$^a$}Dipartimento di Fisica,
Universit\`a di L'Aquila, I-67010 Coppito, AQ, and\\
INFN, Laboratori Nazionali del Gran Sasso,
I-67010 Assergi, AQ, Italy \\
E-mail: \email{gianfagna@aquila.infn.it}, 
        \email{fabrizio.nesti@aquila.infn.it}\\[1ex]
\noindent\llap{$^b$}Sezione INFN di Ferrara, I 44100 FE\\
E-mail: \email{giannotti@fe.infn.it}}

\abstract{A modification of the relation between axion mass and the PQ
constant permits a relaxation of the astrophysical constraints,
considerably enlarging the allowed axion parameter space. We develop this idea 
in this paper, discussing a model for an {\it ultramassive} axion, which 
essentially represents a supersymmetric Weinberg-Wilczek axion of the
mirror world. The experimental and astrophysical limits allow a PQ
scale $f_a\sim 10^4-10^6$ GeV and a mass $m_a\sim$ MeV, which can be
accessible for future experiments.

On a phenomenological ground, such an {\it ultramassive} axion
turns out to be quite interesting. It can be produced  
during the gravitational collapse or during the merging of 
two compact objects, and its subsequent decay into $e^+e^-$ 
provides an efficient mechanism for the transfer of 
the gravitational energy of the collapsing system to the 
electron-positron plasma. This could resolve the 
energy budget problem in the Gamma Ray Bursts and also 
help in understanding the SN type II explosion phenomena.
}


\begin{document}

\section{Introduction}

The strong CP problem is one of the most puzzling points of modern
particle physics (for a general reference see e.g.~\cite{kim}). 
It resides in the presence of the so-called
$\theta-$term in the QCD Lagrangian, 
$\mathcal{L}_{\theta}=\theta(\alpha_s/8 \pi)G_{\mu\nu}\tilde G^{\mu\nu}$,  
where $\alpha_s$ is the fine structure constant of the strong
interactions and $G_{\mu\nu}$ is the gluon field strength tensor.
The {\it $\theta$-term}, which receives a contribution also from
the complex phases in the quark mass matrices $M_{U,D}$ so that
its effective value is $\ov{\theta}=\theta+\arg(\det \rm
M_U \det \rm M_D)$, is CP violating and leads to a neutron
electric dipole moment $d_{n}$, experimentally not observed. This
implies a very strong upper limit on the parameter $\ov\theta$,
$|\ov{\theta}|<10^{-9}$, which has no theoretical explanation in the
context of the Standard Model.

In the most appealing solution of this problem, the Peccei-Quinn
(PQ) mechanism~\cite{PQ}, the $\theta$ parameter\footnote{In the following 
we will use $\theta$ instead of $\ov{\theta}$ for simplicity.} becomes a
dynamical field, the axion $a=f_a\theta$, and emerges as the
pseudo-Goldstone mode of a spontaneously broken global axial
symmetry $U(1)_{\rm PQ}$. Here $f_{a}$ is a constant, with
dimension of energy, also called {\it axion decay constant}. We
will use the following convention throughout this
paper~\cite{raffelt}: we indicate with $f_{\rm PQ}$ the VEV of a
scalar (or a VEV of a combination of several scalars) responsible for
the $U(1)_{\rm PQ}$ symmetry breaking. The constant $f_a$, which
characterizes axion phenomenology, is defined as $f_{\rm PQ}/N$,
where $N$ stands for the color anomaly of $U(1)_{\rm PQ}$
current.\footnote{ The PQ charges are normalized so that each of
the standard fermion families contributes as $N=1$. 
Therefore, in the Weinberg-Wilczek (WW) model~\cite{WW} 
we have $N=N_g$, where $N_g(=3)$ is the number of fermion families. 
The same holds in the Dine-Fischler-Srednicki-Zhitnitskii (DFSZ) 
model~\cite{DFSZ}.
Other models of the invisible axion, e.g.\ the hadronic 
axion~\cite{KSVZ} or archion~\cite{archion}, generally contain some
exotic fermions and so $N\neq N_g$.}

At quantum level, the PQ symmetry is broken by the chiral anomaly, and since this is a
dynamical effect, its strength is measured by non-perturbative QCD contributions. 
Therefore all the axion properties are essentially related only to the 
PQ scale $f_{\rm PQ}$ and the QCD scale $\Lambda$.

On a phenomenological ground, all the axion characteristics can be
roughly estimated from the pion properties. Indeed, axions
generically mix with pions so that their mass, as well as their
couplings with photons and nucleons, are roughly given by
$f_{\pi}/f_a$ times those of the $\pi$-meson,  
where $f_{\pi}\approx 93$ MeV is the pion decay constant. 
So it is clear that the PQ constant is the relevant scale for axion
phenomenology. For example, in the most general context, axion
interaction with fermions and photons can be described by the
following Lagrangian terms:
\begin{equation}  \label{a-couplings}
\mathcal{L}_a = ic_{ae}\frac{m_e}{f_{a}}\,a \,\bar e\gamma_5
e+ic_{aN}\frac{m_N}{f_{a}}\,a\, \bar N\gamma_5 N + c_{a\gamma}
\frac{\alpha}{8 \pi f_{a}} \,a\, F_{\mu\nu}\tilde F^{\mu\nu} +
\cdots\, ,
\end{equation}
where $m_k$ represents the fermion mass (e.g.\ $m_e,m_N,...$ for
electrons, nucleons, etc.), $\alpha$ is the fine structure constant
and $c_{ai}$ are constant coefficients. The factors $c_{aN}$, 
which refer to axion-nucleon interaction, $N=p,n$, 
are generally of order one,
while for the axion-electron and axion-photon interaction the
related coefficients $c_{ae}$, $c_{a\gamma}$ are model dependent.

As for the axion mass, in general it can be obtained from 
the expression~\cite{Choi}:
\begin{equation}  \label{massa}
m_a^2 = \frac{1}{f_a^2} \frac{VK}{V + K\mathrm{Tr}M^{-1} }\; ,
\end{equation}
where $V=\langle \bar{q}q\rangle \sim \Lambda^3$ is the light
quark condensate responsible for the chiral symmetry breaking,
$M= \mathrm{diag}(m_u,m_d,...)$ is the mass matrix of light quarks
(with $m_q < \Lambda$) and 
$K \sim \Lambda^4$ accounts for the strength of the instanton
induced potential. Then, by taking into acount only two light quarks, 
$u$ and $d$, and using the relation 
$(m_u + m_d)\langle \bar{q}q\rangle = m_\pi^2 f_\pi^2$, 
from (\ref{massa}) one directly arrives to the more
familiar formula:
\begin{equation}  \label{a-m}
m_a = \frac{1}{f_a} \left(\frac{m_u m_dV}{m_u+m_d}\right)^{1/2} =
\frac{z^{1/2}}{1+z}\cdot \frac{f_\pi m_\pi}{f_a} \approx
\left(\frac{10^6~\mathrm{GeV}}{f_a}\right) \times 6.2 ~\mathrm{eV},
\end{equation}
where $z=m_u/m_d\simeq 0.57$.

In the  WW model~\cite{WW} the PQ symmetry is broken by two Higgs
doublets $H_{1,2}$ with the VEVs $v_{1,2}$, namely $f_a =
(v/2)\sin 2\beta$, where $v=(v_1^2 + v_2^2)^{1/2}\simeq 247$ GeV
is the electroweak scale and $\tan\beta=v_2/v_1$. Therefore, the
WW axion is quite heavy, and its mass can vary from a few hundred
keV to several MeV:
\begin{equation}  \label{WW-mass}
m_a^{\mathrm{WW}} = \frac{2N}{v\sin 2\beta} \left(\frac{m_u
V}{1+z}\right)^{1/2} \approx \frac{150 ~\mathrm{keV} }{\sin
2\beta}.
\end{equation}
However, its couplings with fermions and photons are too strong and
for this reason the WW model is completely ruled out for any values
of the parameter $\beta$ by a variety of terrestrial
experiments such as the search of the decay $K^+\to \pi^+ a$, the
$J/\psi$ and $\Upsilon$ decays into $a+\gamma$, the nuclear
deexcitations via axion emission, the reactor and beam dump
experiments, etc.~\cite{PDG}.

For any realistic axion model, these experimental constraints 
generally imply the lower bound on the PQ scale, 
$f_a \gapprox 10^4$ GeV, which in turn implies $m_a\lapprox 1$ keV. 
This is what happens in all the so-called invisible axion models.
Anyway, such a light axion can easily be produced inside a star at
a temperature of a few keV, and can accelerate the cooling process
in a dangerous way. So the axion is required to interact only
weakly with fermions in order to strongly suppress the energy
transport process inside the star. Also, the axion luminosity from
the SN core must be constrained in order to not ruin the neutrino
signal detected at the time of SN 1987A explosion~\cite{raffelt}. 
These astrophysical considerations exclude all scales $f_a$ up to
$10^{10}$ GeV~\cite{kim,raffelt}, and so $m_a < 10^{-4}$
eV.\footnote{ In the case of the hadronic axion~\cite{KSVZ} or
archion~\cite{archion}, a small window around $f_a \sim 10^6$ GeV
(axion mass of a few eV) can be also permitted. } On the
other hand, the cosmological limits related to the primordial
oscillations of the axion field or to the non-thermal axion
production by cosmic strings, demand the upper bound $f_a \lapprox
10^{11}-10^{12}$ GeV~\cite{kim,raffelt}. Thus, not much parameter space
remains available. 

Is it possible to relax the astrophysical constraints? 
It is quite interesting to note that all astrophysical constraints 
from stellar evolution could be satisfied for the PQ scales 
above the laboratory limit, $f_a > 10^4$ GeV, up to values 
order $10^7$ GeV, if the axion mass would remain in the MeV 
range, as with the mass of the WW axion (\ref{WW-mass}). 
However, the tight relation between the axion mass 
$m_a$ and the PQ scale $f_a$ (\ref{a-m}) is very constraining 
and does not allow such a situation. 

If there were, in fact, another source for axion mass, 
e.g.\ from Plank scale induced effects, then
the axion could change its properties and then no longer be useful for the
solution of the strong CP problem. This explains why this relation
is universally accepted, and in several papers the astrophysical and
cosmological limits are given in terms of axion mass instead of the PQ
constant. 

But is the relation (\ref{a-m}) really universal? What would happen, in
fact, if the axion could communicate with another, {\it hidden},
sector of particles and interactions? In general, 
if the axion potential were to get dominant contribution from the 
hidden sector, then we would expect it to solve the strong CP problem 
for that sector rather than for our observable world. 
However, this is not necessarily the case. 

In particular, one can assume that the hidden sector is 
a mirror world, a parallel sector of "mirror" particles and interactions  
with the Lagrangian 
completely similar to that of the ordinary particles~\cite{mirror}. 
In other words, it has the same gauge group and coupling constants 
as the ordinary sector, 
so that the Lagrangian of the whole theory is invariant with respect 
to the Mirror parity (M-parity) under the interchange of the two sectors. 
Several phenomenological and astrophysical implications of the 
mirror world have been studied in the literature~\cite{mirror1}. 
In particlular, it could provide a new insight for the cosmological 
dark matter made up by mirror baryons~\cite{B-L}. 
For a recent review of the mirror matter concept one can refer 
to~\cite{M-review}. 

We can further assume that the ordinary and mirror sectors have 
the common Peccei-Quinn symmetry~\cite{BGG}, with the same PQ charges 
carried by the ordinary Higgs doublets $H_{1,2}$ and their 
mirror partners $H'_{1,2}$. 

If the M-parity is an exact symmetry, then the particle physics should
be exactly the same in two worlds, and so the strong CP problems would
be simultaneously solved in both sectors.  In particular, the axion
would couple to both sectors in the same way and their
non-perturbative QCD dynamics would produce the same contribution to
the axion effective potential.  This situation would not bring
drastic changes of the axion properties; just the axion mass would
increase by factor of $\sqrt2$ with respect to the standard expression
(\ref{a-m}).\footnote{However, even in this less interesting case
there would be significant modification of the primordial oscillations
of axion field and their contribution to dark matter of the
Universe~\cite{BGG2}.}

However, a very interesting situation emerges if the M-parity is
spontaneously broken in the Higgs 
sector, so that the mirror electroweak scale $v'$ is
considerably larger than the ordinary one 
$v$~\cite{BM,BDM}.\footnote{Cosmological implications of such a
scenario were studied in refs.~\cite{BDM,asym}.}  In this case, as it
was shown in ref.~\cite{BDM,BGG}, also the mirror QCD scale $\Lambda'$
becomes larger than the ordinary one $\Lambda$ and thus one expects
that the dominant contribution to the axion potential comes from the
mirror sector.  On the other hand, if the latter contribution is
predominant, the axion mass relation with the PQ scale could change in
a considerable way. Of course, in this case it is absolutely not
evident that the axion can still solve the strong CP problem. Indeed,
if we ask for it to be dominantly governed by the mirror QCD, it is
natural to expect that it will cancel the {\it mirror} and not the
ordinary $\theta$-term. Fortunately, this is not the case: this axion
can still solve our strong CP problem, as far as we ask for the Yukawa
structure to be the same in the two sectors~\cite{BGG}.  So it appears
that the mirror extension of the standard axion model is the only way
that allows for a quite heavy axion without spoiling the PQ mechanism.

In particular in this paper we will consider an axion with the mass $m_a > 1$ MeV
and with a PQ scale $f_a \sim 10^6$ GeV. Certainly such a
massive axion cannot ruin stellar evolution process, even though it
can still be produced in the hot SN core at a temperature of a few
10 MeV.
We expect then that
the available parameter range is sizeably increased with respect
to the standard DFSZ axion model. 

As we will show, such an axion can have many interesting
astrophysical implications. In particular, it can help in 
understanding the Gamma Ray Bursts (GRBs) and the supernova 
explosion phenomena via the mechanism suggested in 
ref.~\cite{axidragon}\footnote{We underline that the possible relation 
between GRBs and supernova was
contemplated in the past. In \cite{lisbona}, e.g., is considered
the possibility that the emission of a light axion from SN can provide the energy
necessary for the GRBs production.}.
In fact, such a non standard axion can be
produced during the merging of two compact objects, and then,
due to its decay, 
efficiently transfer the gravitational energy of the collapsing
system into an ultrarelativistic $e^+e^-$ plasma, the fireball, far
from the impact place. Also such a heavy axion, produced in the
core of a SN type II, can decay into $e^+\;e^-$ before reaching
the stellar surface, transferring in this way a huge amount of
energy at a distance of about $1000$ Km from the stellar
core, helping the SN explosion (thermal bomb). On the other hand,
because of the different size of SN type Ib/c, some axions will be
able to leave their surface and then decay into photons,
explaining the observed events of weak GRBs related to the SN type I.

This logic was essentially suggested in a previous paper~\cite{BGG},
but the allowed mass was not enough to open a new window for the axion
of interest in astrophysics. In fact, the parameter space
in~\cite{BGG} does not allow for an $f_{a}>$ few $10^4$ GeV. Observe
that the QCD scale in the mirror sector (and , as a consequence, the
axion mass) is fixed by the renormalization group (RG) evolution of
the strong coupling constant, and so it depends on the matter content
of the theory, as well as on the mirror fermion masses. This
observation suggests to us that a supersymmetric version of the model
can lead to a quite different phenomenology. The results obtained
in~\cite{BGG} are summarized in figure~2 of the cited paper. We
observe that the described axion is constrained, from SN data, in the
region between the dot-dashed line and the dashed one. Also, from
terrestrial experiments, the PQ constant $f_a$ must be bigger than
$10^4$ GeV if the axion mass stays below $\sim$ 1 MeV (twice the
electron mass), otherwise the limit on $f_a$ should be increased to
$f_a\gapprox$ few $10^5$ GeV. We see that for $m_a>1$ MeV, the raising
of the dashed line is strongly suppressed. This is because, for an
axion heavier than two electrons, the total axion decay width
$\Gamma_{tot}$ is dominated by the axion decay into electrons and
positrons. Thus the photon flux is strongly suppressed. This phenomena
can be improved if the axion mass is increased for a given PQ constant
$f_a$, since $\Gamma(a\rightarrow e^+ e^-)\sim m_a/f_a^2$. This, as we
will show, is exactly what happens in a supersymmetric model, for a 
certain range of $f_a$. 
We will then consider a supersymmetric mirror axion
model and this will allow us to strongly enlarge the parameter space
for the axion with relevant phenomenological consequences.

%
Finally, in this model the hierarchy problem between the mirror and the
ordinary e.w.\ scale is solved naturally via the GIFT 
(Goldstones Instead of Fine Tuning) mechanism~\cite{GIFT1}. 
In fact, the Higgs potential
has an additional $SU(4)$ symmetry. This {\it accidental} symmetry
is global, but contains the local one $SU(2)\times SU(2)\equiv SO(4)$. 
When the M-parity is broken, the mirror Higgses acquire a large VEV, while
the ordinary ones remain as Goldstone bosons until
the supersymmetry breaking is taken into account. 
The ordinary Higgs VEVs are then 
naturally of the order of the supersymmetry soft breaking scale~\cite{GIFT1,GIFT2}. 

The paper is organized as follows.
In sect.~2, we present a simple mirror supersymmetric axion model and
describe the general features and the experimental and astrophysical
bounds, showing that there is an allowed parameter space which can be
of interest for the future experimental search. In sect.~3, we
carefully study the interesting relations with the physics of the GRBs
and with the dynamics of different supernovae. Finally, in sect.~4, we
summarize our results.

\section{The Mirror Weinberg-Wilczek axion}

Let us consider a model based on the gauge symmetry $G\times
G^{\prime}$ where $G=SU(3)\times SU(2)\times U(1)$ stands for the
standard model of the ordinary particles: the quark and lepton fields
$\psi_i =q_i,~l_i, ~u^c_i, ~d^c_i, ~e^c_i$ ($i$ is a family index) and
two Higgs doublets $H_{1,2}$, while $G^{\prime}=SU(3)^\prime\times
SU(2)^\prime \times U(1)^{\prime}$ is its mirror gauge counterpart
with the analogous particle content: the fermions $\psi^{\prime}_i=
q^{\prime}_i,~l^{\prime}_i, ~u^{c\prime}_i, ~d^{c\prime}_i,
~e^{c\prime}_i$ and the Higgses $H^{\prime}_{1,2}$~\cite{mirror}. From
now on, all fields and quantities of the mirror sector have an apex to
distinguish them from the ones belonging to the ordinary world. All
fermion fields $\psi,\psi^{\prime}$ are taken in a left-chiral basis.

Let us assume that the theory is invariant under the mirror parity M:
$G\leftrightarrow G^{\prime}$, which interchanges all corresponding
representations of $G$ and $G^{\prime}$. Therefore, the two sectors
are described by identical Lagrangians and all coupling constants
(gauge, Yukawa, Higgs) have the same pattern in both of them. In
particular, for the Yukawa couplings
\begin{eqnarray}  \label{SM-Yuk}
&& \mathcal{L}_{\mathrm{Yuk}} = G_U^{ij} u^c_i q_j H_2 + G_D^{ij}
d^c_i q_j H_1 +
G_E^{ij} e^c_i l_j H_1 + {\rm h.c.}, \nonumber \\
&& \mathcal{L}^{\prime}_{\mathrm{Yuk}} = G_U^{\prime ij}
u^{c\prime}_i q^{\prime}_j H^{\prime}_2 + G_D^{\prime
ij}d^{c\prime}_i q^{\prime}_j H^{\prime}_1 + G_E^{\prime
ij}e^{c\prime}_i l^{\prime}_j H^{\prime}_1 +{\rm h.c.},
\end{eqnarray}
we have $G_{U,D,E}^{ij} = G^{\prime ij}_{U,D,E}$. In addition, the
initial $\theta$-terms are equal, $\theta = \theta^\prime$.

We further assume that the two sectors have a common Peccei-Quinn
symmetry $U(1)_{\rm PQ}$ realized {\it \`a la} Weiberg-Wilczek
model. The essential point is then to have a term in the Higgs
potential which mixes the Higgses of different kinds. The simplest
possibility is:
\begin{equation}  \label{pot-mix}
\mathcal{V}_{\mathrm{mix}} = -\kappa (H_1 H_2)^\dagger
(H^{\prime}_1 H^{\prime}_2) ~ + ~ \mathrm{h.c.},
\end{equation}
where the coupling constant $\kappa$ should be real due to
M-parity.\footnote{Notice that in the limit $\kappa =0$ there
emerge two separate axial global symmetries, $U(1)_A$ for the ordinary
sector under which $\psi_i \to \exp(-i\omega/2)\psi_i$ and
$H_{1,2} \to \exp(i\omega) H_{1,2}$,
 and $U(1)^{\prime}_A$ for the mirror sector:
$\psi^{\prime}_i \to \exp(-i\omega^{\prime}/2)\psi^{\prime}_i$ and
$H^{\prime}_{1,2} \to \exp(i\omega^{\prime}) H^{\prime}_{1,2}$.
Therefore, the term $\mathcal{V}_{\mathrm{mix}}$ demands that
$\omega^{\prime}=\omega$ and thus it reduces $U(1)_A\times
U(1)^{\prime}_A$ to its diagonal subgroup $U(1)_{\rm PQ}$.}

As far as the M-parity is an exact symmetry, the particle physics
should be exactly the same in the two worlds: in particular, the
quark mass matrices are identical, $M_{U,D}=M^{\prime}_{U,D}$, the
QCD scales coincide, $\Lambda =\Lambda^{\prime}$, and the axion
couples with both QCD sectors in the same way: $f_a^{-1}a(G\tilde{G} +
G^\prime\tilde{G}^\prime)$, so that their non-perturbative
dynamics should produce the same contributions to the axion
effective potential. Clearly, in this case, the strong CP problem
is simultaneously solved in both worlds -- the axion VEV cancels
the $\theta$-terms both in the mirror and ordinary sectors.
In such a realization, however, $f_a$ remains order $100$ GeV and
thus it is excluded on the same phenomenological grounds as the
original WW model.

\FIGURE[t]{
\includegraphics[width=12cm,angle=0]{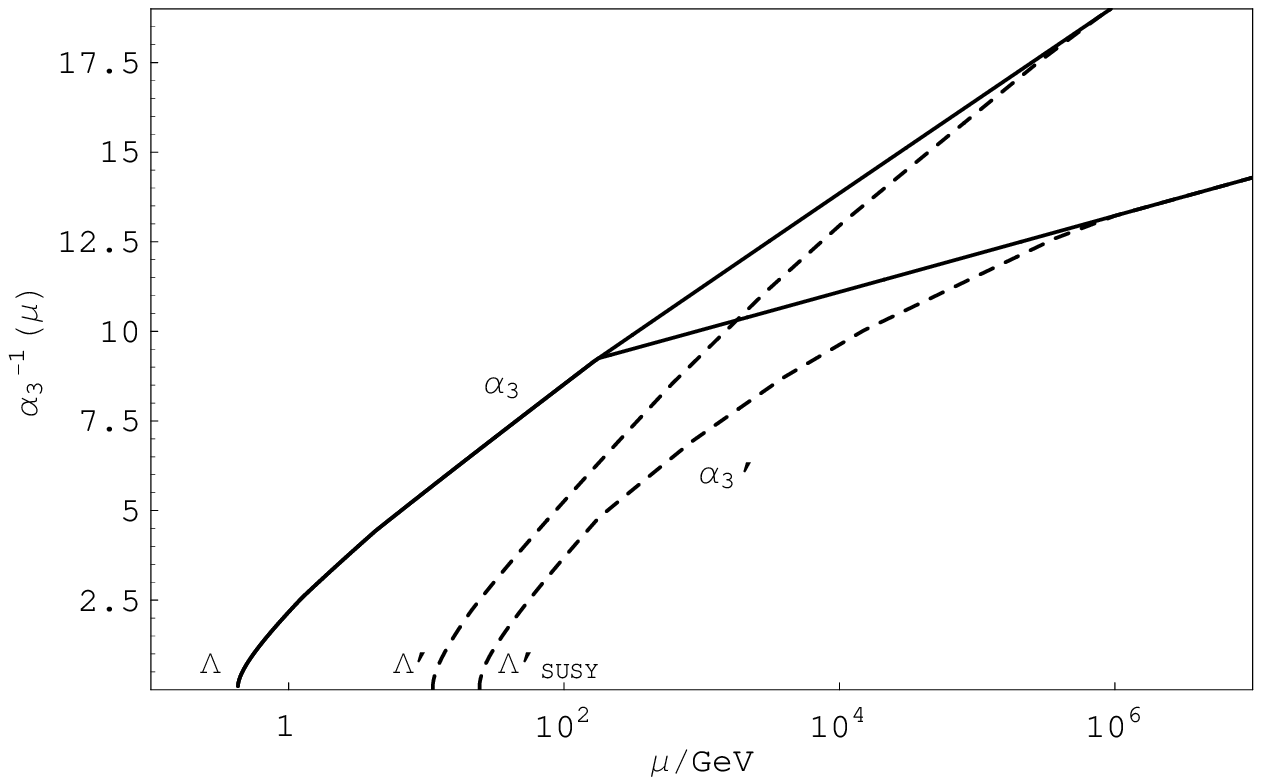}
\caption{Example of the renormalization group evolution of the strong coupling
constants $\alpha_s$ and $\alpha^{\prime}_s$ (respectively solid and
dashed), as a function of the energy scale $\mu$, for $v^{\prime}/v=
10000$. 
Supersymmetry is supposed to be broken at the
scale $m_s=m_t$.}
\label{fig:RG}
}

The situation is more interesting when the M-parity is
spontaneously broken, as it was suggested in ref.~\cite{BM}, and
the electroweak symmetry breaking scale $v^{\prime}$ in the mirror
sector becomes larger than the ordinary one $v$. Since $U(1)_{\rm
PQ}$ is a common symmetry for the two sectors, the PQ scale is
determined by the larger VEV $v^{\prime}$, i.e.\ $f_{\rm PQ}\sim
(v^{\prime}/2)\sin 2\beta^\prime$. The axion state $a$ dominantly
comes from the mirror Higgs doublets $H^{\prime}_{1,2}$, up to
small ($\sim v/v^\prime$) admixtures from the ordinary Higgses
$H_{1,2}$. Hence, it is a WW-like axion with respect to the mirror
sector, while it couples with the ordinary matter as an invisible
DFSZ-like axion. This would lead to somewhat different particle
physics in the mirror sector and it is not {\it a priori} clear that the
strong CP problem can still be simultaneously fixed in both
sectors. However, as it was shown in~\cite{BGG}, this is just the
case as long as the Yukawa structure in the two sectors is the
same.
It happens then that the mirror quark masses are scaled
linearly
with respect to the ordinary ones: $m^{\prime}_{u,c,t}= \zeta_2
m_{u,c,t}$, $m^{\prime}_{d,s,b}= \zeta_1 m_{d,s,b}$, where
$\zeta_2 = v^\prime_2/v_2=\zeta(\sin\beta^\prime/\sin\beta)$ and
$\zeta_1 = v^\prime_1/v_1=\zeta(\cos\beta^\prime/\cos\beta)$. At
very high energies, $\mu \gg v^{\prime}$, the strong coupling
constants $\alpha_s(\mu)$ and $\alpha^{\prime}_s(\mu)$ should be
equal due to M-parity. Under the renormalization group (RG)
evolution they both evolve down in parallel ways until the energy
reaches the value of mirror-top mass $m^{\prime}_t \simeq \zeta_2
m_t$. Below it, $\alpha^{\prime}_s$ will have a different slope
than $\alpha_s$, and this slope will change every time below the
mirror quark thresholds $m^{\prime}_b \sim \zeta_1 m_b$, etc. In
the evolution of $\alpha_s$ these thresholds occur at lower
scales, $\mu = m_t,m_b$, etc. Then it is very easy to determine
the scale $\Lambda^{\prime}$ at which $\alpha^{\prime}_s$ becomes
large, once we know that for the ordinary QCD this happens at
$\Lambda\simeq 200$ MeV. In other words, $\Lambda'$ becomes a
function of $\zeta_{1,2}$, and for $v^{\prime}\gg v$ one could
obtain a significant difference between the QCD scales:
$\Lambda^{\prime}>\Lambda$ (see figure~\ref{fig:RG}).

The value of the QCD constant in the mirror sector depends
on the RG evolution of the mirror strong coupling. This, in turn,
depends on the matter content of the theory. Therefore in a
supersymmetric theory we expect a different result.
In general, the relation between the ordinary and mirror QCD scale can
be written as
\begin{equation}
\frac{\Lambda^\prime}{\Lambda}= A(\beta,\beta^\prime)\,\zeta^\rho,
\end{equation}
where $A$ is a function of the angles $\beta$ and $\beta^\prime$,
while $\rho$ is a constant.  They both depend on the number of mirror
light quarks.  We can estimate,\footnote{We assume, for simplicity,
that there are no light quarks in the mirror sector. This is easily
verified when $f_a \gapprox$ few $10^4$ GeV.} for a non supersymmetric
theory,
$\rho\simeq 0.36$, while in a supersymmetric theory we
find the bigger value\footnote{For this estimation we have assumed that the supersymmetry
is broken at the top quark mass. It is also assumed that there are no
mirror light quarks, and that the lightest mirror quark is heavier
than the ordinary top quark. These assumptions are verified for
$f_a\gapprox$ few $10^6$ GeV.}
$\rho\simeq 0.54$.  

\bigskip

In the following we will consider the renormalizable Lagrangian described by the
superpotential:
\begin{equation}  \label{W}
\mathcal{W}=\mathcal{T}(\mathcal{YR}-M^2)+\mathcal{S}
(\mathcal{H}_1 \mathcal{H}_2+\mathcal{H}_1^{'} \mathcal{H}_2^{'}
-\mathcal{R}^2),
\end{equation}
where $M$ is an energy scale $M \gg v$,
$\mathcal{H}_1,\mathcal{H}_2, \mathcal{H}_1^{'},\mathcal{H}_2^{'}$
are the Higgs superfields and
$\mathcal{T},\mathcal{Y},\mathcal{R},\mathcal{S},$ are other
superfields. $\mathcal{R}$, $\mathcal{Y}$, $\mathcal{S}$ have
respectively the PQ charges
$\omega_{\mc{R}}=-\omega_{\mc{Y}}=1/2$, $\omega_{\mc{S}}=-1$,
while for the Higgs fields $\omega_{1}=\sin^2 \beta$,
$\omega_{2}=\cos^2 \beta$ and the same for $\omega^\prime_{1}$ and
$\omega^\prime_{2}$ with $\beta$ replaced by $\beta^\prime$.
$\mc{T}$ does not transform under PQ symmetry. For convenience we
indicate with $T,Y,R,S,H_i,H_i^{'}$ the scalar components of the
superfields $\mc{T},\mc{Y},\mc{R},\mc{S},\mc{H}_i,\mc{H}_i^{'}$.
The second term on the right hand side of (\ref{W}) fixes the VEV
pattern of the ordinary and mirror Higgses. The constraint
equation is $\br H_1 H_2\kt +\br H_1^{\prime} H_2^{\prime}\kt \sim
\br R^2 \kt$ where $\br R^2 \kt$ is fixed in the first term on the
right hand side of (\ref{W}): $\br R \kt \sim \br Y \kt \sim M$.
Since the ordinary electroweak scale $\br H \kt$ is of the same
order as the supersymmetry breaking scale $m_s$, the previous
equation fixes $\br H^{\prime}\kt \sim \br R \kt \sim M$.
Therefore the $R$ VEV fixes the M-parity breaking scale of the
theory. Since $M \gg v \sim m_s$ we expect all the mirror fermion
masses to be heavier than the ordinary ones and, as a consequence,
$\Lambda^\prime>\Lambda$.

The complete M-invariant scalar potential consists in the sum of
the F and D terms, plus the supersymmetry breaking contribution:
\begin{equation}  \label{Vtot}
\mathcal{V}= \mathcal{V}_F+\mathcal{V}_D+\mathcal{V}_B.
\end{equation}
More explicitly
\begin{equation}  \label{VS}
\mathcal{V}_F= \tilde{\mathcal{V}}(H,H^{'}) +
|TR|^2+|YR-M^2|^2+|TY-2RS|^2,
\end{equation}
where
\begin{equation}  \label{VH}
\tilde{\mathcal{V}}(H,H^{'})=
|H_1 H_2+H^{'}_1 H^{'}_2-R^2|^2+S^2(H_1^2+ H_2^2+H^{'2}_1+
H^{'2}_2)+h.c,
\end{equation}
while:
\begin{eqnarray}  \label{Dterm}
\mathcal{V}_D&=&\frac{g^2}{2}\left[(H_1^+ H_2)^2 +(H_1^{'+}
H_2^{'})^2 \right]+ \\  \nonumber
&&{}+\frac{g^2+g_y^2}{8}\left[(H_1^+ H_1-H_2^+ H_2)^2+ (H_1^{'+}
H_1^{'}-H_2^{'+} H_2^{'})^2 \right],
\end{eqnarray}
where $g$ is the coupling constant for isospin and $g_y$ refers to the
hypercharge. 
Finally $\mathcal{V}_B$ contains all the possible soft
supersymmetry breaking terms
and fixes the scale of the supersymmetry breaking $m_s$.

In this model the hierarchy problem between the mirror and ordinary 
Higgs VEVs is solved in a rather natural way by the Pseudo-Goldstone 
or GIFT mechanism~\cite{GIFT1}. Namely, while the mirror Higgses 
get VEVs order $M \gg M_W$, the ordinary Higgses can get the masses 
(and hence the VEVs) of the order $m_s$.  

More precisely, in the supersymmetric limit $\mathcal{V}_B=0$, 
the F-term potential (\ref{VH}) has an accidental global 
symmetry $SU(4)$, larger than the local symmetry $SU(2)\times SU(2)'$ 
acting on the Higgses $H_{1,2}$ and $H'_{1,2}$ respectively. 
Therefore, if the non-zero VEVs are located on the mirror Higgses, 
then we obtain $v'_1v'_2 = R^2 \sim M^2$, 
and the D-term (\ref{Dterm}) gives  $\tan\beta'=v'_2/v'_1 = 1$, 
while the ordinary Higgses $H_{1,2}$ remain as Goldstone superfields. 

Then, considering the soft terms $\mathcal{V}_B$, 
one generates the soft mass terms for $H_{1,2}$ and, 
as a remarkable property of the GIFT mechanism~\cite{GIFT1,GIFT2}, 
also supersymmertic $\mu$-term, $\mu H_1H_2$ with $\mu\sim m_s$, 
as far the VEV $\langle S\rangle \sim m_s$ is generated by the 
soft terms $\mathcal{V}_B$.
As a result, one can generate non zero VEVs $v_1,v_2\sim m_s$ 
(with $v_1^2 + v_2^2= v^2 = (247~ \rm GeV)^2$), and 
the parameter $\tan\beta=v_2/v_1$ in general can be different 
from 1. 
On the other hand, given that the mirror VEVs are generated 
at a large scale, $v'_1=v'_2 \sim M$, 
the soft terms $\mathcal{V}_B$ cannot significantly shift 
their values and thus we remain 
with $\tan \beta ^{'}\approx 1$.

\bigskip

Let us discuss now in more detail the axion phenomenology in this
model. The PQ scale is (see~\cite{kim}) $f_{\rm PQ} =\sqrt{\sum_j
(v_{j}\omega_{j})^2}$ where $\omega_{j}$ and $v_{j}$ are
respectively the PQ charges and VEVs of the scalar fields. We find
$ f_{\rm PQ}=(f^2+f^{\prime \;2}+\br S \kt^2+ 1/2 M^2)^{1/2}$ where
$f=(v/2)\sin 2 \beta$ and $f'=(v'/2)\sin 2 \beta^\prime$. In the
interesting physical limit $f^\prime \sim M\gg f\sim m_s$ we see
that $f_{\rm PQ}$ is essentially equal to $f^\prime$.

\begin{figure}[t]
\begin{center}
  \includegraphics*[height=4.5cm,angle=0]{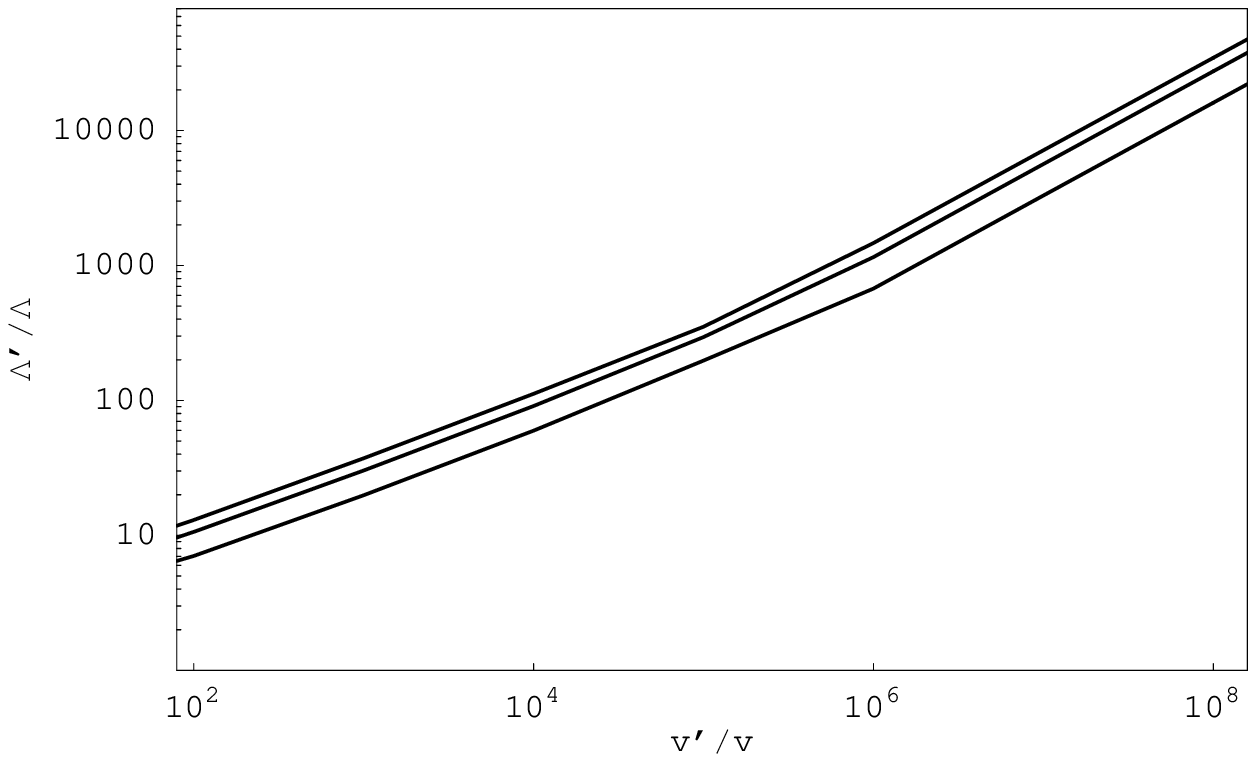}\quad
\includegraphics*[height=4.5cm,angle=0]{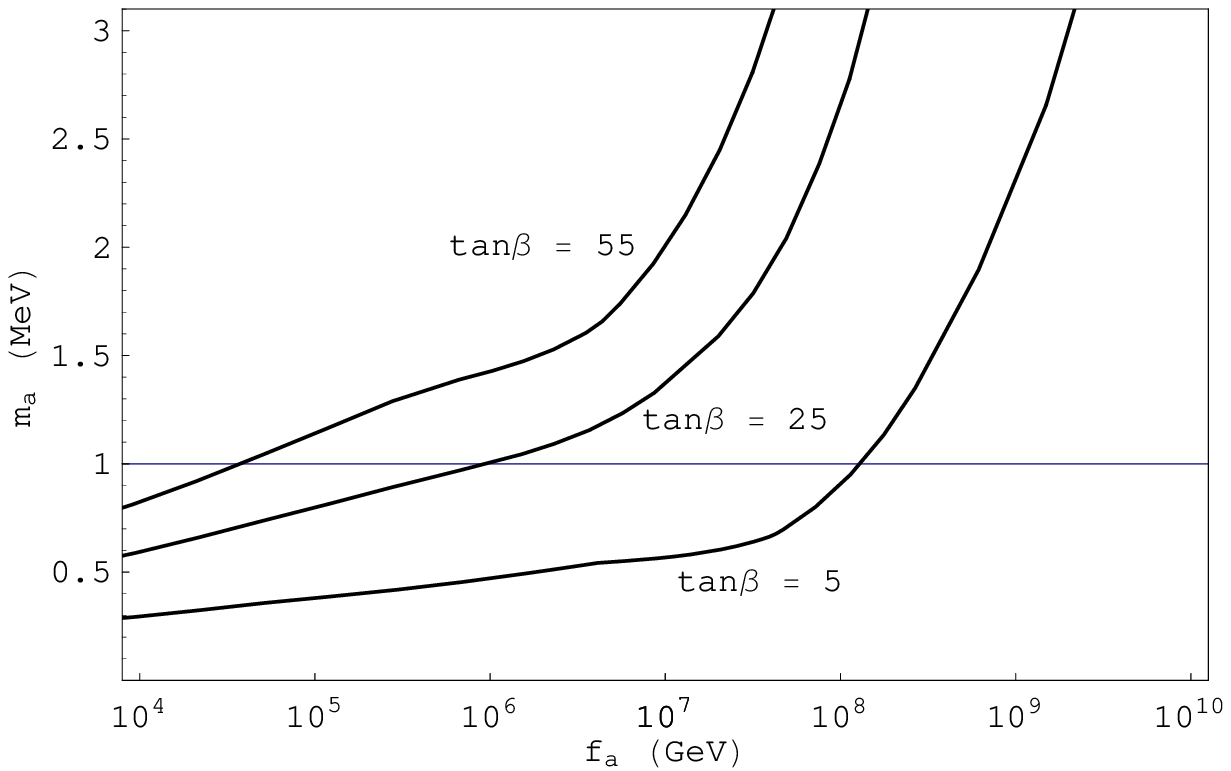}
\caption{Pictorial representation of the mirror
QCD scale and mirror axion mass, in relation to the ordinary
value. All the three lines correspond to $\tan \beta^\prime=1$ and, from the bottom, to $\tan
\beta=5$, $\tan \beta=25$, $\tan \beta=55$. 
Supersymmetry is supposed to be broken at the
scale $m_s=m_t$.}
\end{center}
\label{fig:two}
\end{figure}

For what concerns the axion mass, its square is in general given by
the sum of the right hand side of (\ref{massa}) and a similar
term, but with $V,\,K,\,M$ replaced by
$V^\prime,\,K^\prime,\,M^\prime$. The meaning of these parameters
is obvious, $M^\prime$is the mass matrix of the mirror light
quarks, and the values $K^{\prime}\sim \Lambda^{\prime 4}$ and
$V^{\prime}\sim \Lambda^{\prime 3}$ characterize the mirror gluon
and quark condensates. Assuming, for simplicity, that there are no
light quarks in the mirror sector and that
$\Lambda^{\prime}\gg\Lambda$, we find $m_a\simeq \frac{\;
\,\Lambda^{\prime\,2}}{f_a}$. The axion
mass is then essentially driven by the mirror QCD. This mass can be
several times bigger than the ordinary WW mass:
\begin{equation}\label{massa'}
\frac{m_a}{m_a^{WW}}\simeq
\frac{\sin 2\beta}{C}\left(\frac{\Lambda^\prime}{\Lambda}\right)^2,\qquad
C=2\left(\frac{m_u}{(1+z)\Lambda}\right)^{1/2}\sim 0.2.
\end{equation}
For the numerical computation we have used the following values of
the quark masses: $m_u=4$ MeV, $m_d=7$ MeV, $m_s=150$ MeV (at
$\mu=1$ GeV), and $m_c(m_c)=1.3$ GeV, $m_b=4.3$ GeV and $m_t=170$
GeV (respectively at $\mu=m_c,m_b,m_t$). For the parameters $V$
and $K$, related to the quark and gluon condensates, we have taken
$V = (250 ~\mathrm{MeV})^3$ and $K = (230 ~\mathrm{MeV})^4$ and we
have assumed that the corresponding parameters in the mirror
sector scale as $V^{\prime}/V = (\Lambda^{\prime}/\Lambda)^3$ and
$K^{\prime}/K = (\Lambda^{\prime}/\Lambda)^4$. In figure~\ref{fig:two} is shown
that $\Lambda^\prime$ can be significantly larger than $\Lambda$,
allowing for an axion mass $\sim 1$ MeV when $f_a\sim 10^6-10^7$
GeV.

The axion couples with fermions and photons in the standard way, with a
strength inversely proportional to the PQ constant~\cite{kim}.
Referring to (\ref{a-couplings}) we find:
\begin{equation}  \label{Yuk-eud}
c_{ae} = \frac1N\sin^2\!\beta\,, \quad c_{ad} = \frac1N\sin^2\!\beta-
\frac{1\,z}{1+z}\,, \quad c_{au} = \frac1N\cos^2\!\beta- \frac{1}{1+z}\,.
\end{equation}
For the axion-nucleon couplings we consider $c_{aN}\sim 1$,
while
the axion interaction with photons is measured by:
\begin{equation}  \label{photon}
c_{a\gamma} = \frac83 - \frac{6K \mathrm{Tr}(M^{-1}Q^2)} {V + K
\mathrm{Tr}(M^{-1}) } \simeq \frac{2z}{1+z},
\end{equation}
where the trace is taken over the light quark states ($u,d$), $Q$
are their electric charges $(+2/3,-1/3)$, and $\alpha=1/137$ is
the fine structure constant. In addition, our axion couples with
mirror photons, with the constant:
\begin{equation}  \label{photon-mir}
c^{\prime}_{a\gamma}= \frac83 -
\frac{6K^{\prime}\mathrm{Tr}(M^{\prime -1} Q^{\prime 2})}
{V^{\prime}+ K^{\prime}\mathrm{Tr}(M^{\prime -1})}\,,
\end{equation}
where the factor $C^{\prime}$ for the case of two
($u^\prime,d^\prime$),  one ($u^\prime$) or no light quarks
respectively takes the values $2z^{\prime}/(1+z^{\prime})$, 0 and
$8/3$.
Hence, the axion decay widths into the visible and mirror photons
respectively are:
\begin{equation}  \label{width}
\Gamma(a\to \gamma \gamma)= \frac{g^2_{a \gamma} m_{a}^3}{64 \pi}\,,
\qquad \Gamma(a\to \gamma^{\prime}\gamma^{\prime})= \frac{g^{\prime
2}_{a \gamma}m_{a}^3}{64 \pi}\,,
\end{equation}
where $g_{a\gamma}=c_{a\gamma}(\alpha/2\pi f_{a})$ and similarly for
$g_{a\gamma}^\prime$.

In addition, if $m_a >2m_e$, the axion can decay also into an
electron-positron pair:
\begin{equation}  \label{width-e}
\Gamma(a\to e^+e^-)= \frac{g^2_{ae} m_{a}}{8 \pi} \sqrt{1 -
\frac{4m_e^2}{m_a^2} }\,,
\end{equation}
where $g_{ae}=c_{ae}(m_e/f_{a})$.

\EPSFIGURE{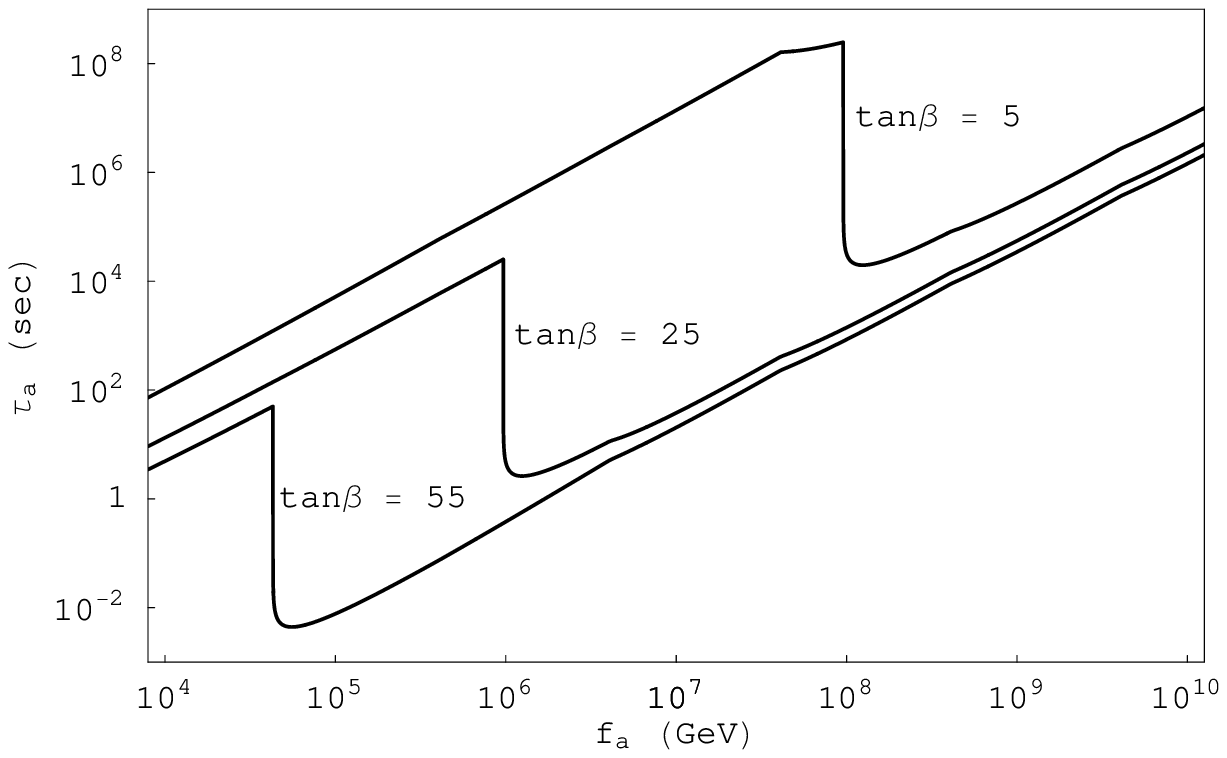,height=5cm}{Axion lifetime as a function of $f_a$ for
  different values of $\tan\beta=5,25,55$. 
  Supersymmetry is supposed to be broken at the
scale $m_s=m_t$.\label{fig:life}}

We present in figure~\ref{fig:life} the axion lifetime in the present
supersymmetric model. Notice that, as soon as the axion mass turns over $1$ MeV
its lifetime is strongly suppressed. In fact in this last case 
its decay width is completely dominated by the $a \to e^+ e^-$ channel. 

For details of the experimental bounds we refer to ref.~\cite{BGG}.
We remark here that the parameter region of interest for us is
$f_a\sim 10^5-10^7$ GeV, a value safe from terrestrial limits.
In this region, the axion mass is about $1$ MeV so it is not constrained by
the standard astrophysical considerations applicable for the DFSZ axion. In fact, since it
is quite heavy, its production rate in the stellar cores, with
typical temperatures $T$ up to 10 keV, is suppressed by the
exponential factor $\exp(-m_a/T)$.
On the other hand, this argument is not applicable to the SN, whose
core temperature is several $10$ keV. If the axion-nucleon couplings
are large enough, $g_{ap},g_{an}> 10^{-7}$ (then $f_a < 10^7$ GeV), the axions are strongly trapped
in the SN, inside a core of radius $R\simeq 10\,$km, and they have
a thermal distribution. In this case the
energy luminosity at $t=1$s can be estimated as
$L_a \simeq f_a^{16/11}\times 3\cdot 10^{50}$ erg/s~\cite{Tur,Bur}.
Hence, for $f_a <$~a~few $10^6$ GeV the axion luminosity $L_a$
becomes smaller than a few $10^{51}$ erg/s, and
the total energy and duration of the SN 1987A neutrino burst
should not be affected.

The constraint related to the axion decay into ordinary photons is
not relevant for an axion with $m_a > 2 m_e$, since then, the total
axion decay width is almost completely driven by $a\to e^+e^-$ 
(see figure~\ref{fig:life}). On
the other hand, another constraint emerges due to axion decay into
mirror photons. Since the ordinary matter is transparent for
the mirror photons, the emission of the latter can lead to the
unacceptably fast cooling of the supernova core. The decay rate
for an axion with energy $E$ into mirror photons is
$(m_a/E)\Gamma^\prime$, where $\Gamma^\prime = \Gamma(a\to
\gamma^\prime \gamma^\prime)$ is given by eq.(\ref{width}).
Therefore, we have $L^{\prime}_\gamma(t) = 4\pi m_a \Gamma^\prime
\int_0^R r^2 n_a(r,t) dr$, where $n_a = 1.2T^3/\pi^2$ is the axion
number density. Taking a core temperature $T$ of about $20-30$
MeV, one can roughly estimate that  $L^{\prime}_\gamma \simeq
\Gamma^\prime m_a (4\pi R^3/3)(1.2T^3/\pi^2)\sim 10^{-2} g^{\prime
2}_{a\gamma} m_a^4$. Hence, the condition $L^{\prime}_\gamma <
10^{51}$ erg/s implies that the decay width $\Gamma^\prime$, for
$m_a \sim 1$ MeV, should not exceed a few s$^{-1}$. In other terms,
using eq.(\ref{photon-mir}), we roughly obtain the bound
$m_a^2/f_a < 10^{-7}$ MeV.

\bigskip

\EPSFIGURE{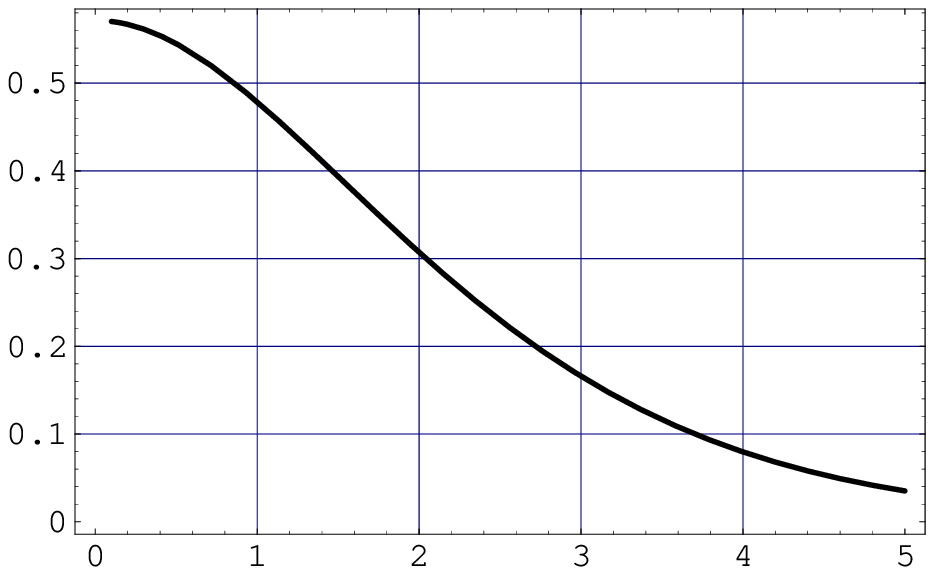,height=4.8cm,angle=0}{Effective extra neutrino
added due to mirror axion as function of axion mass in MeV.
\label{fig:BBN}}

{
\sloppy 

Finally, a possible cosmological problem must be considered. If the
axion contribution to the energy density at the time of the Big
Bang Nucleosynthesis (BBN) is very large, this model can disagree 
with the strongly verified prediction on the
primordial Helium abundance. This energy contribution is expressed
in terms of effective number of extra neutrinos $\delta N_\nu$ in
figure~\ref{fig:BBN}.

At the present, deuterium (D) and $^4 H_e$ data seem to indicate that
a large number of neutrino species is disfavored. There are several
recent papers on this problem (see, e.g.~\cite{fra1} and reference
therein), which give different bounds $\delta N_\nu \leq 0.3$, $\delta
N_\nu \leq 0.5$ or even $\delta N_\nu \leq 1$~\cite{fra1}. These
limits translate to different lower bounds on the axion mass.  

}

\section{Mirror Axion: Supernovae and Gamma Ray Bursts }

The Gamma-Ray Bursts (GRBs) puzzle theorists from many points of
view~\cite{piran}. The most striking feature is that an enormous
energy, up to $10^{53-54}$ erg, is released in a few seconds, in terms
of photons with typical energies of several hundred keV. The
time-structure of the prompt emission and the afterglow observations
well agree with the fireball model~\cite{meszaros} in which the GRB
originates from the $e^+ e^-$ plasma that expands at ultrarelativistic
velocities undergoing internal and external shocks. The Lorentz factor
of the plasma has to be very large, $\Gamma\sim 10^2$, which requires
a very efficient acceleration mechanism. Namely, the $e^+ e^-$ plasma
should not be contaminated by more massive matter (baryons), and
hence the fireball has to be formed in a region of low baryonic
density.

In particular, the fireball could be powered via annihilation
$\nu\bar{\nu}\to e^+ e^-$ of the neutrinos emitted from the dense
and hot medium in the accretion disk around a central black hole
(BH),
which can be formed at the merger of two neutron stars (NS), or a
NS and a black hole (BH)~\cite{ruffert}. One can consider also the
merger of a BH and a white dwarf (WD). In addition, the accretion
disk can be formed by the collapse of a rotating massive star, so
called failed supernova or collapsar~\cite{paczynski}. These
objects could potentially provide the necessary energy budget for
the GRB. Namely, the typical values of the mass $M$ accreted
through a disk, the radial size of a disk $R$ and the accretion
time $t$, can be estimated as:
\begin{eqnarray}\label{budget}
{\rm NS} + {\rm NS}: & ~M\sim 0.1 M_\odot & ~R \sim 50~ {\rm km},
~~ t\sim 0.1 ~{\rm s}
\nonumber \\
{\rm NS} + {\rm BH}: & ~M\sim 0.5 M_\odot & ~R \sim 50~ {\rm km},
~~ t\sim 0.1 ~{\rm s}
\nonumber \\
{\rm Collapsar}: & ~M\sim 2 M_\odot & ~R \sim 200~ {\rm km}, ~~
t\sim 20 ~{\rm s}
\nonumber \\
{\rm WD} + {\rm BH}: & ~M\sim 1 M_\odot & ~ R \sim 10^4~ {\rm km},
~~ t\sim 100 ~{\rm s}.
\end{eqnarray}
However, the problem remains how to transform efficiently enough
the available energy into the powerful GRBs. Due to the low efficiency
of the $\nu\bar\nu \to e^+ e^-$ reaction, the models invoking it
as a source for the GRBs have serious difficulties in reaching such
large photon luminosities. Even though during the collapse of compact
objects an energy  of $\sim 10^{53}\,$erg is normally emitted in terms
of neutrinos, they deposit only a small percent of their
energy to fireball, and take the rest away. In addition, neutrino annihilation
is effective only at small distances, less than $100\,$km, which are
still contaminated by baryon load, and cannot provide a
sufficiently large Lorentz factor~\cite{ruffert}.

In~\cite{axidragon}, it was proposed that the emission of the light
pseudoscalar particles like axions--- which can be effectively
produced inside the accretion disks and then decay into $e^+e^-$
outside the system--- can provide an extremely efficient mechanism
for transferring the gravitational energy of the collapsing system
into the ultrarelativistic $e^+e^-$ plasma. The advantage of using
the decay $a\to e^+e^-$ instead of $\nu\bar\nu\to e^+e^-$
annihilation is obvious. First, it is 100 percent efficient, since
the decaying axions deposit their energy and momentum entirely into
the $e^+e^-$ plasma. And second, the decay can take place in
baryon free zones, at distances of $1000\,$km or more, and so the
plasma can get a Lorentz factor $\Gamma\sim 10^2$.

Here we suggest that the described mirror axions can be the
required pseudoscalar particles. In the dense and hot medium of
the accretion disk it is mainly produced by its bremsstrahlung in
the nucleon-nucleon scattering~\cite{raffelt}. In order to have
efficient production,
its mass should not over-exceed the characteristic temperature of
the matter, typically a few MeV. Assuming for simplicity the
non-degenerate and symmetric baryonic matter, the energy-loss
rate per unit mass due to emission of axions is:
\begin{equation}\label{rate-eps}
\epsilon \simeq g_{N}^2 \rho_{11} T_{\rm MeV}^{7/2} \times 2\cdot
10^{31} ~ {\rm erg}~ {\rm g}^{-1} {\rm s}^{-1},
\end{equation}
where $\rho_{11}$ is the disk density in units of $10^{11}\,$g/cm$^3$ and $T_{\rm MeV}$ is the matter temperature in MeV.

Axions are in the free streaming regime, if their mean free path
is larger than the accretion disk size ($R_{100}=R/100\,$km), which
yields~\cite{axidragon}:
\begin{equation}\label{g-tr}
g_{N} < g_{N}^{tr}(\rho,T) = 2 \times 10^{-6} \rho_{11}^{-1}
T_{\rm MeV}^{1/4} R_{100}^{-1/2}.
\end{equation}
Then the total axion luminosity from the accretion disk with a
mass $M$ can be roughly estimated as:
\begin{equation}\label{lum}
L \simeq \epsilon M \simeq (10^6 g_N)^2 \rho_{11} T_{\rm
MeV}^{7/2} (M/M_\odot)
\times 4\cdot 10^{54} ~ {\rm erg}~{\rm s}^{-1}.
\end{equation}
This value can be so big that the axions can extract all the available
energy from the collapsing system with very high efficiency.

We need the emitted axions to decay into $e^+e^-$ outside the
disk, in the regions of low baryon density,
which correspond to distances of several hundred or thousand km.

For the sake of simplicity, let us fix the parameters as
$m_a =1.5$ MeV, $g_{ae}=10^{-9}$ and $g_{a\gamma}\sim
10^{-12}$ MeV$^{-1}$. In this range of parameters the axion
lifetime can be well approximated by $ \tau \sim \tau(a\to e^+e^-)
= 8\pi g_{ae}^{-2} (m_a^2- 4m_e^2)^{-1/2}$,
since the axion decay width into electrons is much larger than the
one into photons
(cfr.\ref{width},\ref{width-e}).
Therefore, since the mean
decay length of the axion is $D = c\tau E/m_a$, where $E\sim 2T$
is the average energy of the emitted axions, we find:
\begin{equation}\label{D}
D \simeq (10^{9} g_{e})^{-2}
\,m_{\rm MeV}^{-2} \,E_{\rm MeV} \times 5\cdot 10^3 \,{\rm km}\,,
\end{equation}
where $m_{MeV}$ represents $m_a/$MeV and $E_{\rm MeV}=E/$MeV.

In the view of our mechanism, the short GRBs (duration $\sim
10^{-1}$ s), can be naturally explained by the NS-NS merger, with
typical values
$M\sim 0.1M_\odot$ and $t\sim 0.1$ s, the typical density
$\rho_{11}\simeq 1$ and temperature $T\simeq 4$ MeV. Then the
total energy emitted in axions from the accretion torus  can be
roughly estimated as ${\cal E} \simeq Lt \simeq 10^{53}$ erg,
while the mean decay length is $D\simeq 4\cdot 10^3$ km, much
larger than the size of the system ($R\sim 50$ km). Thus, the
axions decay in the baryon clean zones and deposit their energy
entirely to the $e^+e^-$ plasma, which can get a large Lorentz
factor and  give rise to rather isotropic photon emission with
total energies up to $10^{53}$ erg. The relative hardness of the
photon spectrum in the short GRBs well agrees with this situation.
Somewhat more energetic short bursts can be obtained in the case
of the NS-BH merger, with $M\sim 0.5M_\odot$.

\pagebreak[3]

On the other hand, the long GRBs can be related to the collapsar.
In this case, we need to estimate the fraction of energy deposited
from the accretion disk with say $M\sim 2M_\odot$, $R\sim 200$ km
and $t\sim 20$, within a cone of the solid angle $\Omega$ along
the polar axis, where the baryon density is lower and the
funneling of the plasma in this direction can take place producing
a jet expanding outwards. By taking $\rho \sim 10^{10}$ g/cm$^3$
and $T\sim 2$ MeV, we obtain
the beamed GRB with the energy release ${\cal E}/\Omega \sim Lt
\sim 10^{54}$ erg.
As far as axions decay at large distances, about 1000 km, the
Lorentz factor can approach large values, $\Gamma \sim 10^2$. This
analysis is supported by the result of the simulation in~\cite{aloy},
which shows that if the energy would be transferred to the plasma
at distances $\sim 600$ km, a successful burst could be obtained
with $\Gamma \sim 40$.

The axions can be produced also at the supernova explosion. For
$g_{N} \sim 10^{-6}$ they are in the trapping regime in the
collapsing core and are emitted from the axiosphere having a
thermal spectrum with a temperature $T$ of a few MeV~\cite{Bur}.
Therefore, in total, an energy of a few $\times 10^{51}$ erg can be
emitted during the collapse period and subsequent cooling of the
proto-neutron star, in terms of axions with the mean energy $E\sim
3T$. The latter undergo the decay into $e^+e^-$ at the distance
$D\sim 10^3$ km. In this case, the impact of the axion emission
crucially depends on the geometrical size of the collapsing star.

In particular, supernovae type Ib/c result from the core collapse
of relatively small stars, where the hydrogen and perhaps also the
helium shells are missing. Their radius can be as small as $R\sim
10^4$ km, comparable to the axion decay length $D$. This in turn
implies that $\exp (-R/D_a)$ is not very small, and it can be of
order $10^{-3}$ to 1, in which case a reasonable amount of axions
can decay outside the mantle producing a fireball.
So the weaker GRBs associated with a supernova type Ib/c could
take place, having typical energies up to a few $10^{51}$ erg.

The SN type II are associated with large stars, having an extended
hydrogen shell ($R > 10^7$ km). Thus, the axion decay essentially takes
place completely inside the mantle -- the fraction of axions
decaying outside the star, $\exp(-R/D)$, is essentially zero and
thus no GRB can be observed. Indeed, the SN 1987A event did not
show any $\gamma$ signal. On the other hand, the energy of a few
$10^{51}$ erg released by axion decay at distances $\sim 1000$ km
can help to solve the painful problem of mantle ejection (in the
prompt mechanism, shock usually stalls at a distance of a few
hundreds km).

Concluding, the axion emission from the collapsing systems and their
subsequent conversion into the relativistic plasma via the decay $a\to
e^+e^-$ outside these systems could naturally explain a variety of the
GRBs. This mechanism suffers no energy deficit and it makes more
natural the possibility of the plasma acceleration. In particular, the
short GRBs, with timescale $\sim 0.1$ s and total energies up to a few
$\times 10^{53}$ erg can originate from the NS-NS or NS-BH mergers,
while the collapsar could give rise to the longer GRBs, with $t\sim
10-30$ s and ${\cal E}/\Omega$ up to a few $10^{54}$ erg. The events
with $t\sim 100$ s could also be initiated by the BH-WD merger. In
later cases, one can expect more baryon dirty fireball.  All this
well agrees with the observational features of the GRBs.
As explained, an interesting possibility is the association of some weak GRBs
(with total energy up to $10^{51}$ erg or so) with the supernovae
type Ib,c. Considering that the axion emission could also help the
supernovae type II explosion, we see that this model can provide a unified
theoretical base for the GRB and SN phenomena. Interestingly, the
emission of these axions can also be important for explaining the
observed GRB's preceded by supernova explosions, via collapse of
usual neutron stars to quark or hybrid stars~\cite{ber}.

\section{Conclusions}

We have presented a new model of axion for the solution of the
strong CP problem.
The main feature of this model is the modification of the relation between 
the axion mass and the PQ constant, with quite interesting phenomenological
consequences.

We have hypothesized the existence of another sector
of particles and interactions, the {\it mirror world}, which is an
exact copy of our world but with a larger electroweak
scale $v'\gg v$. This difference also implies
a different dynamics in the QCD sector. The fermion masses
are in fact driven by the Higgs VEVs, and the different thresholds
for the mirror masses lead to a pole  $\Lambda^\prime$ (in the
evolution of the mirror strong coupling constant), that can be significantly
higher than the ordinary one $\Lambda$. Since the value of
$\Lambda$ in the two sectors depends on the RG evolution
equations, and consequently on the matter content of the theory,
this behavior can be improved in a supersymmetric
model.

The result is that the relation of axion mass to the PQ constant is
relaxed, and our particle can be quite heavy, maintaining the weak
coupling with matter and photons typical of the invisible axions.
This behavior has great benefit on a phenomenological and
astrophysical ground. Because of its large mass our axion has in
fact no influence on stellar evolution. All astrophysical
constraints come from SN explosion and allow its mass to be
$m_a\gapprox$ MeV with a PQ constant of order $10^5-10^6$ GeV.

Such a heavy axion happens to have quite an interesting
phenomenology, in particular in relation to the GRB and SN
physics. The axion emission from the collapsing systems, and their
subsequent conversion into the relativistic plasma via the decay
$a\to e^+e^-$ outside these systems, could naturally explain the
fireball formation and, consequently, a variety of the GRBs. This
mechanism was first proposed using the reaction $\nu\bar\nu\to
e^+e^-$, instead of the axion decay $a\to e^+e^-$, but the
advantage of using the last decay, instead of the neutrino
annihilation, is clear. First, it is 100 percent efficient,
since the decaying axions deposit their energy and momentum
entirely into the $e^+e^-$ plasma. And second, the decay can take
place in baryon-free zones, at distances of 1000 km or more, so
the plasma can get a Lorentz factor $\Gamma\sim 10^2$.

Another interesting possibility is the association of some weak
GRBs (with total energy up to $10^{51}$ erg or so) with the
supernovae type Ib,c. The axion mean-free path is a few $10^3$ Km,
so a few of them can reach the  supernovae type I surface
and then decay into photons, giving rise to the observed weak
GRBs. Finally, the axion emission could also help the supernovae
type II explosion, thus providing a unified theoretical base for
the GRB and SN phenomena.

As a final note, the parameter window allowed for our axion is
accessible to the axion search in the future reactor and beam dump
experiments.

\vspace{3mm}
{\large \textbf{Acknowledgements}}
\vspace{3mm}

We are grateful to Zurab Berezhiani for useful discussions and technical
help and to Elizabeth Price for the careful reading of this paper.


\begin{thebibliography}{99}

\bibitem{kim}  J.E. Kim, Phys. Rep. 150 (1987) 1;
H.Y. Cheng, \textit{ibid.} 158 (1988) 1.

\bibitem{PQ}
R.D. Peccei and H.R. Quinn, Phys. Rev. D 16 (1977) 1791.

\bibitem{WW}
S. Weinberg, Phys. Rev. Lett. 40 (1978) 223; F. Wilczek,
\textit{ibid.} 40 (1978) 279.

\bibitem{DFSZ}  A.R. Zhitnitskii, Sov. J. Nucl. Phys. 31 (1980) 260;
M. Dine, W. Fischler, M. Srednicki, Phys. Lett. B104 (1981) 199.

\bibitem{KSVZ}  J.E. Kim, Phys. Rev. Lett. 43 (1979) 103;
M. Shifman, A. Vainshtein, V. Zakharov, Nucl. Phys. B166 (1980) 493.

\bibitem{archion}  Z. Berezhiani, Phys. Lett. B129 (1983) 99;
Phys. Lett. B150 (1985) 177; \\
A. Anselm, Z. Berezhiani, Phys. Lett. B162 (1985) 349; \\
Z. Berezhiani, M. Khlopov, Z. Phys. C 49 (1991) 73; Sov. J.
Nucl. Phys. 51 (1990) 739;  51 (1990) 935; \\
Z.G. Berezhiani, M.Yu. Khlopov, R.R. Khomeriki,
Sov.J.Nucl.Phys. 52 (1990) 65; \textit{ibid.} 52 (1990) 344.

\bibitem{Choi}
K. Choi, K. Kang, J.E. Kim, Phys. Lett. B181 (1986) 145.

\bibitem{PDG}  Particle Data Group, Eur. Phys. J. C15 (2000) 1.

\bibitem{raffelt}
G.G. Raffelt, Phys. Rep. 198 (1990) 1.

\bibitem{mirror}
T.D. Li, C.N. Yang, Phys. Rev. 104 (1956) 254; \\
Y. Kobzarev, L. Okun, I. Pomeranchuk, Yad. Fiz. 3 (1966) 1154; \\
M. Pavsic, Int. J. Theor. Phys. 9 (1974) 229; \\
R. Foot, H. Lew, R. Volkas, Phys. Lett. B272 (1991) 67. \\


\bibitem{mirror1}
B. Holdom, Phys. Lett. B166 (1985) 196; \\
S.L. Glashow, Phys. Lett. B167 (1986) 35; \\
E.D. Carlson, S.L. Glashow, Phys. Lett. B193 (1987) 168; \\ 
R. Foot, R. Volkas, Phys. Rev. D 52 (1995) 6595; 
Z. Silagadze, Phys. At. Nucl. 60 (1997) 272; \\
S. Blinnikov, astro-ph/9801015, astro-ph/9902305; \\
R. Foot, Z.K. Silagadze, astro-ph/0404515; \\
Z. Berezhiani, Phys. Lett. B417 (1998) 287.  

\bibitem{B-L}
Z. Berezhiani, D. Comelli, F. Villante, Phys. Lett. B503 (2001) 362; \\
L. Bento, Z. Berezhiani, Phys. Rev. Lett. 87 (2001) 231304; 
Fortsch. Phys. 50 (2002) 489;  hep-ph/0111116; \\
A.Y. Ignatyev, R.R. Volkas, Phys. Rev. D 68 (2003) 023518; \\
R. Foot, R.R. Volkas, Phys. Rev. D 69 (2004) 123510; \\
Z. Berezhiani, P. Ciarcelluti, D. Comelli, F. Villante, 
astro-ph/0312605; \\
P. Ciarcelluti, Ph.D. Thesis, astro-ph/0312607; \\ 
For earlier works see  
S. Blinnikov, M. Khlopov, Astron. Zh. 60 (1983) 632; \\
M. Khlopov {\it et al.}, Astron. Zh. 68 (1991) 42. 
 
\bibitem{M-review} 
Z. Berezhiani, hep-ph/0312335; \\
R. Foot, astro-ph/0407623. 

\bibitem{BGG}
Z. Berezhiani, L. Gianfagna, M. Giannotti, Phys. Lett. B500 (2001)
286; see also V.A. Rubakov, JETP Lett. 65 (1997) 621.

\bibitem{BGG2}
Z. Berezhiani, L. Gianfagna, M. Giannotti, in preparation. 

\bibitem{BM}
Z. Berezhiani, R.N. Mohapatra, Phys. Rev. D 52 (1995) 6607;
see also E. Akhmedov, Z. Berezhiani, G. Senjanovi\'c,
Phys. Rev. Lett. 69 (1992) 3013.

\bibitem{BDM} 
Z. Berezhiani, A.D. Dolgov, R.N. Mohapatra,  
Phys. Lett. B375 (1996) 26. 

\bibitem{asym} 
Z. Berezhiani, Acta Phys. Polon. B27 (1996) 1503; \\
R.N. Mohapatra, V. Teplitz, Astrophys. J. 478 (1997) 29; 
Phys. Rev. D 62 (2000) 063596; \\ 
V. Berezinsky, A. Vilenkin, Phys. Rev. D 62 (2000) 083512; \\ 
R.N. Mohapatra, S. Nussinov, V.L. Teplitz, Phys. Rev. D 66 (2002) 
063002; {\it ibid} D 68 (2003) 023518. 


\bibitem{axidragon}
Z. Berezhiani, A. Drago, Phys. Lett. B473 (2000) 281.

\bibitem{lisbona}
A. Loeb, Phys.Rev.D48 (1993) 3419;\\ 
O. Bertolami, Astropart.Phys.11 (1999) 357; astro-ph/9901184.

\bibitem{GIFT1} 
K. Inoue, A. Kakuto, H. Takano, Prog. Theor. Phys. 75 (1986) 664; \\
A. Anselm, A. Johansen, Phys. Lett. B200 (1988) 331; \\
A. Anselm, Sov. Phys. JETP 67 (1988) 663; \\
Z. Berezhiani, G. Dvali, Bull. Lebedev Phys. Inst. 5 (1989) 55.  

\bibitem{GIFT2} 
R. Barbieri, G. Dvali, M. Moretti, Phys. Lett. B312 (1993) 137; \\ 
R. Barbieri, G. Dvali, A. Strumia, Nucl. Phys. B432 (1994) 49; \\ 
G. Dvali, Q. Shafi, Phys. Lett. B326 (1994) 238; \\ 
Z. Berezhiani, C. Csaki, L. Randal, Nucl. Phys. B444 (1995) 61; \\
Z. Berezhiani, Phys. Lett. B355 (1995) 481;  hep-ph/9412372;  
hep-ph/9703426; \\   
G. Dvali, S. Pokorski, Phys. Rev. Lett. 78 (1997) 807. 

\bibitem{Tur}  M.S. Turner, Phys. Rev. Lett. 60 (1988) 1797.

\bibitem{Bur}
A. Burrows, M.T. Ressell, M.S. Turner, Phys. Rev. D 42 (1990)
3297.

\bibitem{fra1} F. L. Villante, A. D. Dolgov, hep-ph/0310138.

\bibitem{piran}
T. Piran, Phys. Rep. 314 (1999) 575.

\bibitem{meszaros}
P. Meszaros and M.J. Rees, Astrophys. J. 476 (1997) 261; \\
M. Vietri, Astrophys. J. 488 (1997) L105; \\
E. Waxmann, Astrophys. J. 485 (1997) L5.

\bibitem{ruffert}
M. Ruffert and H.-Th. Janka,
astro-ph/9804132.

\bibitem{paczynski}
B. Paczy\`nski, Astrophys. J. 494 (1998) L45.

\bibitem{aloy}
M.A. Aloy {\it et al.}, astro-ph/9911098.

\bibitem{ber}
Z. Berezhiani {\it et al.}, Astrophys. J. 586 (2003) 1250, astro-ph/0209257 ;\\
Z. Berezhiani {\it et al.}, Nucl. Phys. proc. suppl 113 (2002)
268.

\end{thebibliography}
\end{document}